\documentclass[prb,aps,amsmath,preprint]{revtex4}

\usepackage{graphicx}

\usepackage{dcolumn}

\usepackage{bm}

\begin{document}

\title{High-pressure structural, elastic and electronic \\
properties of the scintillator host material, KMgF$_3$ } 
 
\author{G. Vaitheeswaran$^{1,*}$, V. Kanchana$^{1}$, R. S. Kumar$^{2}$, \\ 
A. L. Cornelius$^{2}$, M. F. Nicol$^{2}$, A. Svane$^{3}$, A. Delin$^{1}$ and B. Johansson$^{1,4}$} 
\affiliation{$^{1}$Applied Materials Physics, Department of Materials Science and Engineering, 
Royal Institute of Technology, Brinellv\"agen 23, 100 44 Stockholm, Sweden \\
$^{2}$High Pressure Science and Engineering Center and Department of Physics, University of Nevada, 
Las Vegas, Nevada 89154, USA \\
$^{3}$Department of Physics and Astronomy, University of Aarhus, DK-8000 Aarhus C, Denmark\\
$^{4}$Condensed Matter Theory Group, Department of Physics, Uppsala University, Box.530, SE-751 21, Uppsala, Sweden}
\date{\today}

\begin{abstract}
The high-pressure structural behaviour of the flouroperovskite KMgF$_3$  
is investigated by theory and experiment. 
Density functional calculations were performed within the local density approximation and
the generalized gradient
approximation for exchange and correlation effects, as implemented within
the full-potential linear muffin-tin orbital method. 
{\it In situ} high-pressure powder x-ray diffraction experiments were
performed up to a maximum pressure of 40 GPa using synchrotron 
radiation. 
We find that the cubic $Pm\bar{3}m$ crystal symmetry persists throughout the pressure range studied.
The calculated ground state properties  -- the equilibrium lattice constant, bulk modulus 
and elastic constants -- are in good agreement with experimental results. By 
analyzing the ratio between the bulk and shear modulii, we conclude that KMgF$_3$ is brittle in nature.
Under ambient conditions, KMgF$_3$ is found to be an indirect gap insulator 
with the gap increasing under pressure.
\end{abstract}

\maketitle

\section {Introduction} 
KMgF$_3$ is a technologically important flouroperovskite.
For example, it is used as a vacuum-ultraviolet-transparent (VUV-transparent) material for lenses in 
optical lithography steppers\cite{Nishimatsu} and in 
electro-optical applications\cite{Paus,Fakuda}. When doped with lanthanide ions,
it is a very promising material for scintillators\cite{dorenbos2000}
and radiation dosimeters\cite{Getkin, Furetta}. 
In addition the physical properties of KMgF$_3$ 
may have implications for understanding of the Earth's lower mantle\cite{Bovin, Price}. 

KMgF$_3$ was first synthesized by van Arkel\cite{van} and has 
a simple cubic perovskite structure at room temperature\cite{Remy}. 
KMgF$_3$ demonstrates great stability 
under high compression and 
has not been found to undergo any phase 
transition at any temperature or pressure, suggesting it may be used as an internal 
X-ray calibrant\cite{Wood}. 
%

Several experimental studies of the ground state properties of KMgF$_3$ have been
performed. The elastic 
constants at ambient pressure have been 
measured by Rosenberg and Wigmore \cite{Rosenberg} and by Reshchikova,\cite{Reshchikova} 
while Jones investigated
their pressure and temperature dependence\cite{Jones}.
%

From the theoretical side,  
electronic structure calculations for KMgF$_3$ have been
carried out by means of linear combination of atomic orbitals\cite{Richard}, including
the effects of doping with transition metal impurities in KMgF$_3$\cite{Richard2}. 
The electronic structures
of divalent 3$d$ transition metal impurities
doped in KMgF$_3$ have been investigated by the pseudopotential method\cite{Kawazoe}, 
and the properties of vacancies were studied by Hartree-Fock cluster calculations\cite{Huang}.
The structural, electronic and 
optical properties of KMgF$_3$ were recently investigated by the 
full-potential linear augmented plane wave (FP-LAPW) method\cite{Khenata}.

The present work is a combined theoretical and experimental study of 
the ground state and high-pressure properties of KMgF$_3$. 
We present the equation of state resulting from high-pressure 
diamond-anvil cell experiments on KMgF$_3$ up to 40 GPa.
We also present 
the equation of state, the elastic constants 
and the electronic structure
from theoretical calculations
using two different approximations for the exchange-correlation functional. 

The remainder of the paper is organized 
as follows. Details of the computational 
method as well as details of the experimental setup are 
outlined in section 2. The measured and calculated equations of state are presented
in section 3 together with calculated ground state properties
and elastic properties.
The electronic structure 
and the pressure variation of the band gap are discussed in section 4.  
Finally, conclusions are given in section 5.

\section {computational and experimental details}
\subsection{The electronic structure method}
The all-electron full-potential linear muffin
tin orbital (FP-LMTO) method\cite{Savrasov} is used  to
calculate the total energies and 
basic ground state properties of KMgF$_3$   presented here.
In this method, the crystal volume is split   into two
regions: non-overlapping muffin-tin
spheres surrounding each atom and the
interstitial region between the spheres. We used a double $\kappa$
spdf LMTO basis (each radial function within the spheres is matched to a
Hankel function in the interstitial region) to describe the valence bands.
In the calculations we included the 3s, 3p, 4s, 4p, and 3d bases for potassium, the
3s, 2p, 3p, and 3d bases for magnesium, and the 2s and 2p bases for fluorine. 
The exchange correlation potential was 
calculated within the local density approximation (LDA)\cite{Vosko} as well as 
the generalized gradient approximation (GGA) scheme\cite{Perdew}.
The charge density and potential inside the muffin-tin spheres 
were expanded in terms of spherical
harmonics up to $l_{max}$=6, while in the
interstitial region, they were expanded in plane
waves, with 14146 waves (energy up to 156.30 Ry) 
included in the calculation. Total energies
were calculated as a function of volume for a 
(16 16 16) k-mesh containing 165 
k-points in the irreducible wedge of the Brillouin zone and 
were fitted to the
Birch equation of state\cite{Birch} to obtain the ground state properties.

The elastic constants were obtained from the variation of the total energy under 
volume-conserving strains, as outlined in Refs.  \onlinecite{Oxides}
and \onlinecite{srcl2}.

\subsection{Experimental details}
The high-pressure x-ray diffraction measurements used a sample of
polycrystalline KMgF$_3$ prepared by the solid state reaction method from 
high purity constituent materials as described elsewhere in several reports\cite{zhao,smith,Chadwick}.
Diffraction patterns collected at ambient 
temperature and pressure showed a cubic ($Pm\bar{3}m$) symmetry with a 
cell parameter a = 4.0060(2) \AA\ for KMgF$_3$ which closely agrees with earlier reports\cite{Muradyan,Ross}. 
High pressures were generated by a Merrill-Bassett type diamond-anvil cell (DAC). 
An 185-$\mu$m sample chamber was formed in a 
rhenium metal gasket with a pre-indention of 60-$\mu$m thickness. The powder 
sample was loaded in the gasket with a few 
ruby grains and silicone fluid as pressure transmitting medium\cite{Shen}.
Diffraction experiments were performed at the 16ID-B undulator beam 
line of the High Pressure Collaborative Team (HPCAT) of the Advanced Photon Source (APS). A 
monochromatic x-ray beam with a wavelength of 0.4218 \AA\ was focused down to a 
size of 30 x 30 $\mu$m. Diffraction images were collected with an image plate 
detector for an exposure time of 10 sec. The distance between the sample and the detector and 
the inclination angle of the image plate were calibrated using a CeO$_2$ standard.

The two dimensional images were subsequently integrated to one 
dimensional diffraction patterns using the Fit2D software\cite{Hammersley}. The 
cell parameters were obtained by analyzing the diffraction patterns with the JADE 
software package and the P-V data obtained was fitted with a second-order
Birch-Murnaghan\cite{Birch} equation of state. The 
standard ruby fluorescence technique
and the newly proposed ruby pressure scale of 
Holzapfel\cite{Holzapfel} were used to obtain the pressures in the sample chamber. 

\begin{figure}
\label{figXRD}
\begin{center}
\includegraphics[width=100mm,angle=0,clip]{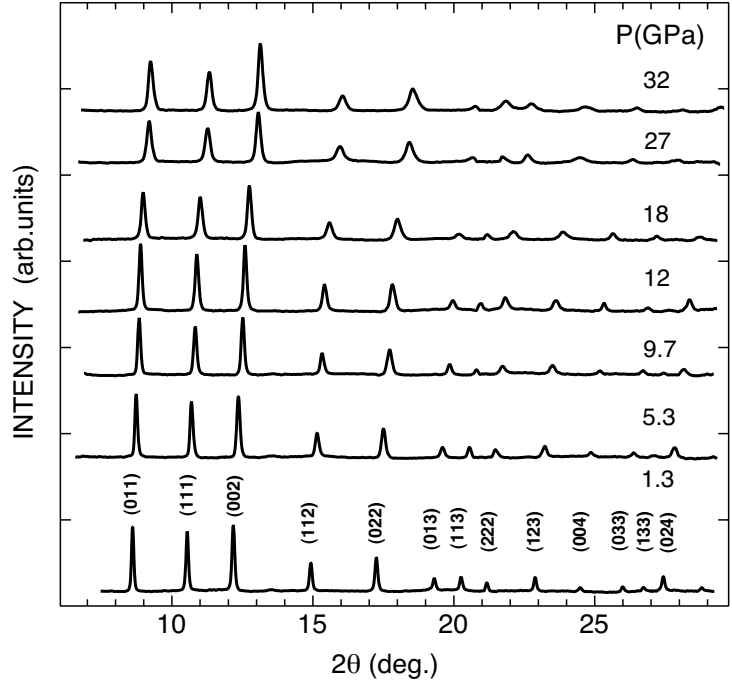}\\
\caption{ Powder x-ray diffraction patterns recorded at various pressures up to
 32 GPa. The indexing in terms of the simple cubic structure is given.}
\end{center}
\end{figure}

\section{Ground state and Elastic properties}
Powder x-ray diffraction patterns collected at several pressures are shown in Figure 1.
On compression, the diffraction patterns remain unchanged up to 40 GPa, 
except for the shifts of diffraction lines caused by the decreasing lattice constant. 
This implies that no structural transformations occur up to 40 GPa in KMgF$_3$.
Figure 2 shows the measured equation of state of 
KMgF$_3$ and compares it with theoretical curves calculated within the 
LDA and GGA. The better theoretical description is obtained with the LDA which is 
somewhat surprising,
since usually the GGA provides an improvement over LDA. At low pressures the
LDA volume is slightly smaller than the experimental one, while  the situation reverses at high
pressures, {\it i. e.,} altogether the LDA predicts KMgF$_3$ 
to be  stiffer than experimental observations. GGA on the
other hand greatly overestimates the equilibrium volume at ambient pressure, which is the main
reason for the poor agreement with experiment. If the GGA curve is scaled througout the pressure
range with the error in equilibrium volume at $P=0$, 
nearly perfect agreement is found with experiment (not
shown).

%
The lattice constant and bulk modulus measured in the present work as well as values
calculated within the LDA and GGA approximations
are given in Table I.
Results from earlier experimental and theoretical works are quoted for comparison.
The bulk modulus obtained in our experiments $B_0$ = 71.2(2) GPa with $B_0'$ = 4.7(3) compares well with 
other experimental results listed in 
Table 1 and also with NaMgF$_3$ reported recently by Liu et al\cite{Liu} ($B=76.0(1.1)$ GPa).
The lattice constant
obtained within the LDA is 1.1 \% lower than the experimental value,
while the corresponding bulk modulus is
22\% higher than the experimental value, which is the usual kind of accuracy 
of LDA. However the calculated LDA lattice constant from the present
work agrees quite well with the experimental work when compared to the earlier FP-LAPW(LDA)
 calculations in which
the reported lattice constant is 2.4\% lower than the experimental value\cite{Khenata}.
The LDA bulk modulus obtained from the
present calculation agrees quite well the published FP-LAPW(LDA) results. When comparing the results
obtained within GGA, the lattice constant is 1.9 \% higher than the experimental value,
 whereas our results for the bulk modulus is within the spread
of the experimental data.
This truly excellent agreement regarding the bulk modulus is, however, a bit fortuitous. 
Since the calculated equilibrium volume is overestimated with GGA (and underestimated 
with LDA), an error -- solely depending on the error in volume -- is introduced in 
the calculated bulk modulus. Therefore, we
recalculated the bulk modulus also at the experimental volume in a manner similar to our earlier 
work\cite{Anna} (see Table I). We find that this diminishes 
the discrepancies between the LDA and GGA results, as expected. In addition, the LDA
bulk modulus now becomes {\it smaller}  than the GGA one for KMgF$_3$, and both functionals are
seen to actually overestimate the bulk modulus, by approximately 14 \% (LDA) and 34 \% (GGA).

The present experiments on KMgF$_3$ relate to recent experiments performed for NaMgF$_3$ and alloys
of NaMgF$_3$ and KMgF$_3$.
The crystal chemistry of Na$_{1-x}$K$_x$MgF$_3$ and NaMgF$_3$ was  studied in detail 
at ambient and at high pressures by Zhao et al. \cite{zhao, zhao2}.
NaMgF$_3$ undergoes a reversible phase transition from orthorhombic ($Pbnm$) to tetragonal 
($P4/mbm$) and then to cubic structure ($Pm\bar{3}m$) upon compression. These phase 
transitions require either compositional changes, by increasing the K concentration to 40\%, or 
changing temperature or pressure. The structural changes in these perovskites are due to
octahedral tiltings and shortening of Mg-F bonds compared to the cubic phase. 
A direct transformation from orthorhombic to cubic structure in NaMgF$_3$,  
however, requires a very high temperature (1038 K).
Moreover, the transition temperature is reported to increase with pressure. 
The temperature dependence of the crystal structure of KMgF$_3$ was recently 
investigated by neutron powder diffraction by Wood et al.\cite{Wood} 
from 4.2 K to 1223 K, and the cubic symmetry was found 
to be stable throughout this temperature range. 
The thermal expansion as well as the atomic displacement parameters 
obtained in their 
experiments show that the F ions behave less
anistropically than in NaMgF$_3$ at such high temperatures.
On comparing these results with the present high-pressure diffraction experiments on KMgF$_3$, 
one may speculate that application of pressure alone would not be sufficient to 
induce structural changes in KMgF$_3$ as 
the cubic phase is very stable.  Such a phase transformation if any, would require either 
application of very high temperature 
or a composition change in the system to achieve changes in the 
order parameters.
Asbrink et al.\cite{Asbrink} have studied single crystals of the transition metal 
bearing perovskite KMnF$_3$, which is isostrucural to KMgF$_3$, under high 
pressure and observed a cubic-to-tetragonal phase transition at a critical 
pressure of 3.1 GPa. On combining these results, phase transitions from 
the cubic symmetry may be expected with a combination of composition change, temperature 
and pressure in KMgF$_3$. A systematic study on the octahedral tilting and order parameters with 
other dopant compositions and the effect of external thermodynamical variables 
are further required to understand the phase stability of KMgF$_3$. 

\begin{figure}
\label{EOS}
\begin{center}
\includegraphics[width=100mm,angle=0,clip]{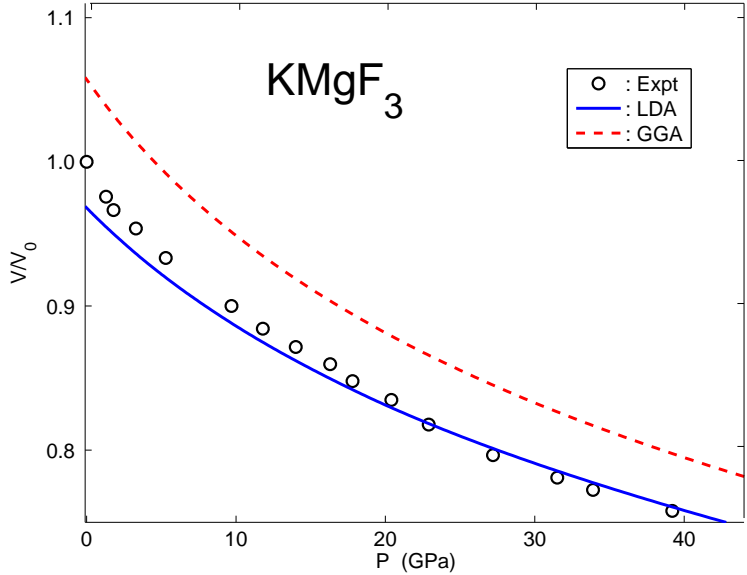}\\
\caption{(Color Online) Equation of state of KMgF$_3$ in the pressure range from 0-40 GPa. The
experimental datapoints are marked by circles, while theoretical results
obtained with LDA and GGA are shown as full (blue) and dashed (red) 
curves. Volumes are given relative to the experimental equilibrium
volume $V_0=64.288$ \AA.
}
\end{center}
\end{figure}


The elastic constants of KMgF$_3$ calculated within LDA and GGA are 
listed in Table  2 where they are also
compared to experimental results as well as earlier calculations. The LDA overestimates all of the 
$C_{11}$, $C_{12}$ and $C_{44}$ elastic constants by between 10\% and  22\%  compared to 
experiment.\cite{Rosenberg,Reshchikova,Jones}
%
The elastic constants obtained within GGA are much closer to 
the experimental values than are the LDA results. 
For instance, both $C_{11}$ and $C_{12}$ are within the
experimental spread. Of course, the elastic constants also 
depend sensitively on the volume, and,  therefore, the same argument as for the bulk modulus can be
applied here. We have, however, refrained from recalculating all the elastic constants with 
the volume correction, but wish to mention that the
excellent agreement between experiment and the GGA elastic constants should be interpreted with care.
Another point of caution is the fact that the calculated values pertain to 0 kelvin, 
while experiments are performed at room temperature. Finite 
temperature generally tends to reduce the elastic constants because of thermal expansion. 
Using the calculated elastic constants we calculated 
the anisotropy factor $A = 2C_{44}/(C_{11} - C_{12})$.
We find an $A =   0.91$ for LDA and $A=1.12$ for GGA. The experimental result is 1.05, 
measured at room temperature\cite{Rosenberg, Reshchikova, Jones}, which is closer to but lower than the
GGA value. However,
the anisotropy factor is found to decrease as the temperature is lowered\cite{Reshchikova}.

A simple relationship, which empirically links the plastic properties of 
materials with their elastic moduli was proposed by Pugh\cite{Pugh}. 
The shear modulus $G$ represents the resistance to plastic deformation, while the bulk modulus $B$ 
represents the resistance to fracture. A high $B/G$ ratio is associated 
with ductility whereas a low value corresponds to 
brittle nature. The critical value which separates ductile and brittle materials is around $1.75$, i.e.
if $B/G > 1.75$ the material behaves in a 
ductile manner, otherwise the material behaves in 
a brittle manner. Frantsevich\cite{Fran}, in a similar fashion has suggested $B/G \sim 2.67$ 
as the critical value
separating brittle and ductile behavior. In the case of KMgF$_3$ the calculated 
value of $B/G$ is 1.5 within LDA and 1.4 within GGA, hence classifying this 
material as brittle. 

Pettifor\cite{Pettifor} suggested that the angular character of atomic 
bonding in metals and compounds, which also relates 
to the ductility, could be described by the Cauchy pressure $C_{12} - C_{44}$. For 
metallic bonding the Cauchy pressure is typically positive. On the other hand, for directional 
bonding with angular character, the Cauchy pressure is negative, with larger negative pressure 
representing a more directional character. These correlations have been verified for ductile materials such as 
Ni and Al that have typical metallic bonding, as well as for brittle semiconductors such as Si with directional
bonding\cite{Pettifor}. 
In the ionic compound KMgF$_3$, the calculated 
Cauchy pressure is  -10 GPa within LDA and -15 GPa within GGA, in good agreement with the nonmetallic 
characteristics of
KMgF$_3$. 

Table 3 presents sound velocities  as derived from the calculated elastic 
constants\cite{Oxides}. The calculated sound velocities agree quite well with the experiments,
in particular for the GGA values, which is a consequence of the somewhat fortuitous
good agreement between the measured and GGA calculated elastic constants.

\section{Electronic structure}

The calculated electron band structure of KMgF$_3$ is shown in Figure 3 
with the
ensuing density of states in Figure 4.
The valence bands consist of the F 
p bands with a gap of 7.24 eV to the conduction band, which is dominated by K states.
The LDA bands are almost identical, however with a gap of only 6.95 eV. The gap increases
almost linearly with compression, at the rate
\[
V\frac{dE_g}{dV}=-7.1 \mbox{ eV}.\]
The conduction band minimum occurs at the $\Gamma$ point, while
the valence band maximum occurs at the R-point $(1/2,1/2,1/2)\frac{2\pi}{a}$. The 
largest occupied energy level at the M-point $(1/2,1/2,0)\frac{2\pi}{a}$ is
marginally lower (by $\sim 0.02$ eV) than the valence band maximum at R, and it remains lower
throughout the pressure range studied here. 
\begin{figure}
\label{figband}
\begin{center}
\includegraphics[width=100mm,angle=0,clip]{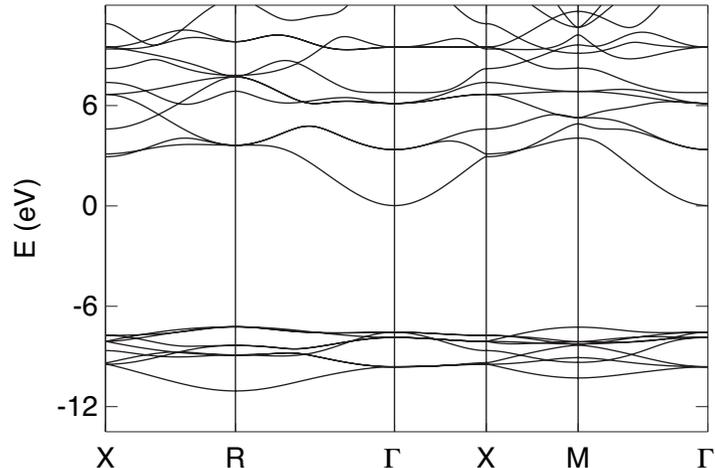}\\
\caption{Band structure of KMgF$_3$ (using GGA, at the experimental lattice constant).
The zero of energy is set at the position of the conduction band minimum.
}
\end{center}
\end{figure}

\begin{figure}
\label{figdos}
\begin{center}
\includegraphics[width=100mm,angle=0,clip]{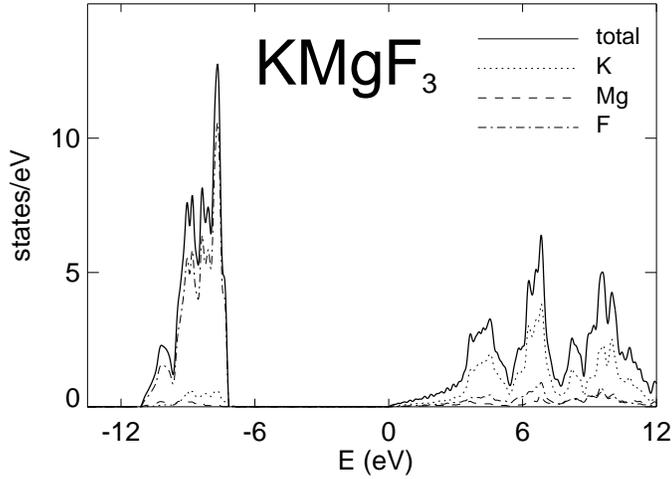}\\
\caption{Density of states of KMgF$_3$ (using GGA, at the experimental lattice constant).
The zero of energy is set at the position of the conduction band minimum.
The partial projections onto the spheres of K, Mg and F are shown with dotted, dashed and
dash-dotted lines, respectively, while full line gives the total density of states. Units
are electrons per eV and per  formula unit.
}
\end{center}
\end{figure}

\section{Conclusions}
In conclusion, we find that pure cubic KMgF$_3$ is very stable under high compression. From our analysis we also find that it is 
a brittle system and an indirect gap insulator whose gap increases with pressure


\acknowledgments
G. V,  V. K, A. D and B. J  acknowledge 
VR and SSF for financial support and SNIC for providing computer time. 
One of the authors (R.S.K) is greatly indebted to his departed colleague 
Yongrong Shen for help in collecting the data. The authors gratefully 
acknowledge the use of the HPCAT facility supported by DOE-BES, DOE-NNSA, NSF, and 
the W. M. Keck Foundation. HPCAT is a high-pressure collaborative access team among the
Carnegie Institution, Lawrence Livermore National Laboratory, the University 
of Nevada Las Vegas, and the Carnegie/DOE Alliance Center. We thank Dr. Maddury Somarazulu and 
other HPCAT staff for technical assistance. This research was supported from the 
U.S. Department of Energy Cooperative Agreement No. DE-FC52-06NA26274 with the University of Nevada Las Vegas.  
\clearpage

\begin{table}[tb]
 \caption{
 Calculated lattice constants (in \AA), Bulk modulus $B_0$ (in GPa)
 and its pressure derivative $B_0'$,
 of KMgF$_3$ at the theoretical equilibrium volume compared with the experiment and other theoretical 
calculations. The bulk moduli have been calculated both at the experimental and theoretical volume
($B_0(V^{\rm{exp}}_{0})$ and $B_0(V^{\rm{th}}_{0})$, respectively)}
\begin{ruledtabular}
\begin{tabular}{cccccc}
& Lattice constant &   $B_0(V^{\rm{th}}_{0})$  & $B_0(V^{\rm{exp}}_{0})$     & $B_0'$  \\
&&&&&\\
\hline
	
       GGA$^a$ & 4.0809     & 72.01   &     97.85       &  4.65  \\
       LDA$^a$ & 3.9630     & 91.47   &     83.23       &  4.79 \\
      LDA, LAPW$^j$   & 3.91      &  90.97  &  -         & 4.64\\
       Expt.   & 4.0060(2)$^a$, 3.973$^b$, 3.978$\pm$0.05$^c$, &       & 71.2(2)$^a$,70.4$^g$,   &  4.7(3)$^a$  \\
               & 3.9897$^d$, 3.993$^e$, 3.9839$^f$ &    & 75.1$^h$,75.6$^i$          \\ 
    
\end{tabular}
\end{ruledtabular}
$^a$Present work, $^b$Ref. \cite{Remy},  $^c$Ref. \cite{Darabont}, 
$^d$Ref. \cite{Ross}, $^e$Ref. \cite{Lee}, $^f$Ref. \cite{David}, 
$^g$Ref. \cite{Rosenberg}, $^h$Ref. \cite{Reshchikova}, $^i$Ref. \cite{Jones}
$^j$Ref. \cite{Khenata}  

\end{table}

\begin{table}[tb]
\caption{
Calculated elastic constants, shear modulus (G), and Young's modulus (E) all expressed in GPa, 
and Poisson's ratio $\nu$
of KMgF$_3$
at the theoretical equilibrium volume}.

\begin{ruledtabular}
\begin{tabular}{cccccccc}
              &$C_{11}$  &$C_{12}$  &$C_{44}$ &  G    &E      &$\nu$  &  \\ \hline

GGA           & 137.0    & 39.5     &  54.6             & 52.3   & 126.3  & 0.208 & Present    \\         
LDA           & 177.0    & 48.7     &  58.7             & 60.9   & 149.5  & 0.228 & Present   \\
LDA           & 119.26   & 38.26     &  63.23            &  -     &  -     &  -   & Ref. \onlinecite{Khenata}  \\
\hline
Expt.         & 132$\pm$1.5  & 39.6$\pm$1.5  &  48.5$\pm$0.6  & - & - & - & Ref.  \onlinecite{Rosenberg} \\
	      & 138$\pm$0.2  & 43.6$\pm$0.2  & 49.83$\pm$0.08 & - & - & - & Ref.  \onlinecite{Reshchikova}  \\
	      & 138.5$\pm$0.5& 44.1$\pm$0.5  &  50.01$\pm$0.1 & - & - & - & Ref.  \onlinecite{Jones} \\
\end{tabular}
\end{ruledtabular}
\end{table}

\begin{table}[tb]
\caption{
Calculated longitudinal, shear, and average wave velocity ($v_l$, $v_s$, and $v_m$, respectively) in m/s
for KMgF$_3$ at the theoretical equilibrium volume}
\begin{ruledtabular}
\begin{tabular}{ccccccc}
             &     &\it{v$_l$}  &\it{v$_s$}  &\it{v$_m$}   \\ \hline

 Present    & LDA & 7402   & 4396 & 4870     \\
            & GGA.& 6706   & 4073  & 4507    \\  
 Expt.      &     & 6470$^a$,6540$^b$  & 3940$^a$,3900$^b$   & 4290$^c$    	    
\end{tabular}
\end{ruledtabular}
$^a$: Wave vector along $<100>$ direction,  Ref.\onlinecite{Rosenberg}, 
$^b$: Wave vector along $<110>$ direction,  Ref.\onlinecite{Rosenberg}, 
$^c$: Ref.\onlinecite{Wood}
\end{table}

\end{document}